\documentstyle[aps,version2,preprint]{revtex}
\begin{document}
\draft
\preprint{OCIP/C 93-15}
\preprint{November 1993}
\begin{title}
Resolved Photon Contributions to\\
Leptoquark Production in $e^+ e^-$ and $e\gamma$ Collisions
\end{title}
\author{Michael A. Doncheski and Stephen Godfrey}
\begin{instit}
Ottawa-Carleton Institute for Physics \\
Department of Physics, Carleton University, Ottawa CANADA, K1S 5B6
\end{instit}

\begin{abstract}
We calculate the resolved photon contribution to leptoquark
production at $e\gamma$ colliders for the center of mass energies
$\sqrt s=500$~GeV and 1~TeV.  We also calculate the resolved photon
contribution to leptoquark production at $e^+ e^-$ colliders for the
center of mass energies $\sqrt{s} = 1$~and~2~TeV.  In both cases we
find that these contributions are considerably larger than the
standard contributions considered in the literature.
\end{abstract}

With the prospect of a high energy linear $e^+ e^-$
collider\cite{ee}, there has been much interest in searching for
physics beyond the Standard Model\cite{BSMee}.  Additionally, there
is a growing interest in the physics that can be accomplished at high
energy $e\gamma$ colliders.  High energy, high luminosity photon
beams can be achieved by backscattering a low energy laser against a
high energy electron beam\cite{egamma}. The luminosity is comparable
to that of the original electron beam with only a small decrease in
energy from the initial electron beam energy.  Some of the physics
that has been explored at $e\gamma$ collisions are measurement of the
$WW\gamma$ coupling\cite{tgv}, Higgs production\cite{higgs},
supersymmetric particle production\cite{susy}, $t$-quark
production\cite{tquark}, and leptoquark
production\cite{leptoLN,HP,BLN,me}.  With this interest, there is the
growing realization that contributions from the hadronic content of
the photon are important and cannot be
neglected\cite{photont,photone}.

Recently, leptoquark production at $e\gamma$ colliders\cite{leptoLN}
and at $e^+ e^-$ colliders\cite{HP,BLN,me} has been studied.
Leptoquarks appear in a large number of extensions of the standard
model such as grand unified theories, technicolour, and composite
models.  In this brief report we examine the resolved photon
contribution to leptoquark production in $e\gamma$ collisions and
find that they are considerable, totally overwhelming the standard,
direct, contributions.

The  most general $SU(3)\times SU(2) \times U(1)$ invariant scalar
leptoquarks satisfying baryon and lepton number conservation have
been written down Buchm\"uller {\sl et al.}\cite{buch}.  However,
only those leptoquarks which couple to electrons can be produced in
$e\gamma$ collisions and for real leptoquark production the chirality
of the coupling is irrelevant.  For this case the number of
leptoquarks reduces to four which can be distinquished by their
electromagnetic charge; $Q_{em}= -1/3$, $-2/3$, $-4/3$, and $-5/3$.
In our calculations we will follow the convention adopted in the
literature\cite{leptoLN} where the leptoquark couplings are replaced
by a generic Yukawa coupling $g$ which is scaled to electromagnetic
strength $g^2/4\pi=\kappa \alpha_{em}$ and allow $\kappa$ to vary.

The process we are considering is shown if Fig. 1.  The parton level
cross section is trivial, given by:
\begin{equation}
\sigma(\hat{s})=\frac{\pi^2 \kappa \alpha_em}{M_s}
                \delta(M_s - \sqrt{\hat{s}}).
\end{equation}
Convoluting the parton level cross section with the quark
distribution in the photon one obtains the expression
\begin{eqnarray}
\sigma(s) & = & \int f_{q/\gamma}(z,M_s^2) \hat{\sigma}(\hat{s}) dz
                \nonumber \\
& = & f_{q/\gamma}(M_s^2/s,M_s^2)
      \frac{\mbox{$2\pi^2\kappa \alpha_{em}$}}{\mbox{$s$}}.
\end{eqnarray}
We would like to point out here that the interaction Lagrangian used
in Ref. \cite{HP} associates a factor $1/\sqrt{2}$ with the
leptoquark-lepton-quark coupling.  Thus, one should compare our
results with $\kappa$ to those in Ref. \cite{HP} with $2\kappa$.

In Fig.~2 the cross sections are shown for $\sqrt{s}=500$~GeV and
1~TeV $e\gamma$ colliders.
The cross section for leptoquarks coupling to the $u$ quark is larger
than those coupling to the $d$ quark.  This is due to the larger $u$
quark content of the photon compared to the $d$ quark content which
can be traced to the larger $Q_q^2$ of the $u$-quark.  For all four
of the leptoquark types we show curves for three different
distributions functions: Drees and Grassie (DG)\cite{drees}, Gl\"uck,
Reya and Vogt (GRV)\cite{GRV}, and Abramowicz, Charchula and Levy
(LAC) set 1\cite{abramowicz}.  Although the cross sections vary, they
are qualitatively the same and agree in order of magnitude.
Comparing the resolved photon contributions to the direct
contributions we find that the former cross sections are roughly two
orders of magnitude larger than the latter.  Clearly they are
important and cannot be ignored.

In Fig. 3 the cross sections are shown for $\sqrt{s} = 1$~TeV and
2~TeV $e^+ e^-$ colliders.  This cross section is obtained by
convoluting the expression for the resolved photon contribution to
$e \gamma$ production of leptoquarks, Eqn. (2), with the
Weizs\"acker-Williams effective photon distribution
\begin{equation}
\sigma(e^+ e^- \rightarrow X S) = \frac{2 \pi^2 \alpha_{em}\kappa}{s}
    \int_{M_s^2/s}^1 \frac{dx}{x} f_{\gamma/e}(x,\sqrt{s}/2)
    f_{q/\gamma}(M_s^2/(x s), M_s^2)
\end{equation}
with the Weizs\"acker-Williams effective photon distribution given by
\begin{equation}
f_{\gamma/e}(x,E) = \frac{\alpha_{em}}{2 \pi} \left\{
   \frac{[1 + (1 - x)^2]}{x} \ln \left[ \frac{E^2}{m_e^2}
     \frac{(1 - 2 x + x^2)}{1 - x + x^2/4)} \right]
   + x \ln \left( \frac{(2 - x)}{x} \right)
   + \frac{2(x - 1)}{x} \right\}.
\end{equation}
As is the case for Fig.~2, the cross section for $Q = -1/3$ and
$-5/3$ leptoquarks is larger due to the larger $u$ quark distribution
in the photon.  We show curves for the four different leptoquark
charges, using the same three resolved photon distribution functions
as in Fig.~2.  Again, although the cross sections vary somewhat, they
are qualitatively in good agreement.  Our results are about an order
of magnitude larger than those of Ref. \cite{HP}.  Ref. \cite{BLN}
also present results for leptoquark production at $e^+ e^-$
colliders.  Those authors calculate the contribution from
$\gamma + e \rightarrow q + S$, regulating the collinear divergence
with the quark mass, which in some ways is similar to including the
resolved photon contribution.  However, this method introduces
uncertainty into the problem (constituent or current quark mass?) and
misses completely the vector meson dominance contribution to the
resolved photon.

There are a number of issues we did not consider in this paper.
Because we wanted to make a direct comparison with previous
calculations of leptoquark production we assumed, as did the previous
calculations, that the photon energy is equal to the beam energy and
did not include the backscattered laser photon spectra.
The main effect of including the
photon spectrum would be to scale the photon spectrum by about a
factor of 0.8.  The second issue to consider is that if one is not
careful there is a double counting in the resolved and direct
contributions to the cross section.  One has to be careful in
matching the low $Q^2$ $p_T$ cut to the resolved photon contributions.
Finally, we did not perform a sophisticated analysis of the
signature, including leptoquark decays, detector acceptances, and
backgrounds expected in a realistic situation.  It is possible that
once these effects are taken into account, the resolved contributions
could be suppressed far more than the unresolved contributions.  The
answer to these questions awaits a more detailed analysis than is
relevant to the main point of this paper.

In conclusion, we have calculated the resolved photon contribution to
single leptoquark production in $e\gamma$ and $e^+e^-$ collisions.
We find that
these contributions are significantly larger than the direct
contributions and should be included in any realistic study of
leptoquark production in $e\gamma$ and $e^+e^-$ collisions.

\acknowledgments

This research was supported in part by the Natural Sciences and
Engineering Research Council of Canada. The authors are grateful to
Manuel Drees and Drew Peterson for helpful communications.

\figure{The resolved photon contribution for leptoquark production in
$e\gamma$ collisions.}

\figure{The cross sections for leptoquark production due to resolved
photon contributions in $e\gamma$ collisions.  (a) is for
$\sqrt{s}=500$ GeV and (b) for $\sqrt{s}=1$~TeV.  In both cases the
solid, dashed, dotdashed line is for resolved photon distribution
functions of Abramowicz, Charchula and Levy \cite{abramowicz},
Gl\"uck, Reya and Vogt \cite{GRV}, Drees and Grassie \cite{drees},
respectively.}

\figure{The cross sections for leptoquark production due to resolved
photon contributions in $e^+ e^-$ collisions.  (a) is for
$\sqrt{s}=1$~TeV and (b) for $\sqrt{s}=2$~TeV.  In both cases the
solid, dashed, dotdashed line is for resolved photon distribution
functions of Abramowicz, Charchula and Levy \cite{abramowicz},
Gl\"uck, Reya and Vogt \cite{GRV}, Drees and Grassie \cite{drees},
respectively.}

\end{document}